\documentstyle[12pt]{article}
\begin{document}
\title{On kinematical constraints in Regge calculus}
\author{V. Khatsymovsky \\
 {\em Institute of Theoretical Physics} \\
 {\em Box 803} \\
 {\em S-751 08 Uppsala, Sweden\thanks{Permanent adress (after 15
November 1993): Budker Institute of Nuclear Physics, Novosibirsk
630090, Russia}} \\

 {\em E-mail address: khatsym@rhea.teorfys.uu.se\thanks{Permanent
E-mail address (after 15 November): khatsym@inp.nsk.su}}}
\date{\setlength{\unitlength}{\baselineskip}
\begin{picture}(0,0)(0,0)
\put(9,13){\makebox(0,0){UUITP-24/93}}
\put(9,12){\makebox(0,0){}}
\end{picture}
}
\maketitle
\begin{abstract}
In the (3+1)D Hamiltonian Regge calculus (one of the coordinates, $
t$, is continuous) conjugate variables are (defined on triangles of
discrete 3D section $ t=const$) finite connections and antisymmetric
area bivectors. The latter, however, are not independent, since
triangles may have common edges. This circumstance can be taken into
account with the help of the set of kinematical (that is, required to
hold by definition of Regge manifold) bilinear constraints on
bivectors. Some of these contain derivatives over $ t$, and taking
them into account with the help of Lagrange multipliers would result
in the new dynamical variables not having analogs in the continuum
theory. It is shown that kinematical constraints with derivatives are
consequences of eqs. of motion for Regge action supplemented with the
rest of these constraints without derivatives and can be omitted; so
the new dynamical variables do not appear.
\end{abstract}
\newpage

The main complication of the physical 4D general relativity (GR) as
compared to it's low-dimensional models is that antisymmetric tensor
bivectors enter theory as natural canonical variables rather than
tetrad of which these depend bilinearly; so some constraints should
be imposed on these bivectors to take into account their tetrad
nature; see, e.g., \cite{Rom}. The same refers to the Regge calculus
version of general relativity (suggested by Regge \cite{Regge}; see
recent review \cite{Will-rev}). Now in the tetrad-connection
formulation \cite{Kha} the 6 6-component bivectors of 2-faces
(triangles) of any 4-tetrahedron at one of it's vertex should be
expressed in terms of the 4 4-component vectors of tetrahedron's
edges beginning at the same vertex. But there is new complication
here connected with that the length of a link calculated from
bivectors in different tetrahedrons sharing this link should be the
same (whereas in the continuum theory tetrads in the different points
are independent). This circumstance can be taken into account, as we
shall see, by writing additional bilinear constraints on bivectors;
when passing to the continuous time form some of these constraints
will contain the time derivatives and thus would spoil canonical
structure of Regge calculus thus far similar to that of continuum GR.

In this note we just show that these imposed by hand constraints with
derivatives assigned to ensure desired kinematical (i.e. not
specifical for gravity) structure of Regge manifold are, in fact,
consequences of the dynamical eqs. of motion for Regge action and of
other constraints without derivatives. Thus, relatively simple form
of kinetic term in Regge calculus is kept.

Now  we briefly discuss Regge Lagrangian of \cite{Kha} in the more
general form and modified notations used in \cite{Kha1} where the
more detailed consideration is given. In order to develop canonical
formalism one should make one of coordinate, call it time $ t$,
continuous by tending dimensions of simplices in some direction to
zero. To perform this limiting procedure in a nonsingular way it is
necessary to issue from some regular arrangement of vertices w.r.t. $
t$-axis. Therefore we consider the sequence of $ t=const$ 3D Regge
manifolds, call these leaves, separated by distances $
O(dt),~dt\rightarrow 0$ such that for each vertex $ i$ in any of the
leaf there are both it's image $ i^+$ and pre-image $ i^-$ at the
distances $ O(dt)$ in the two neighboring leaves. Besides that, we
assume the same scheme of connection of different vertices by links
in the leaves. Then the full 4D Regge manifold is supposed to result
from these 3D leaves by triangulating the space between each two
neighboring ones by new, diagonal and infinitesimal links. By
infinitesimal and diagonal we call  simplices between the two leaves
if their measure (length, area, etc.) is $ O(dt)$ and finite,
respectively; those completely contained in the leaf will be called
leaf simplices. Infinitesimal links connect each vertex with it's
image in the neighboring leave. Diagonal link connects only those
vertices in the neighboring leaves whose images in one of these
leaves are connected by leaf link.

Now, if we denote vertices of the given leaf $ i,~k,~l,...$, the
(3+1)D Regge manifold between each two neighboring leaves proves to
be divided into infinitesimal 4-prisms with bases being just usual
tetrahedrons in the leaf which we denote $ (iklm)$ where round
brackets in such the notations will always mean unordered sequence of
vertices of the given simplex. There are local frames living in
tetrahedrons, connections $ \Omega$ (SO(3,1) matrices connecting
these frames) and antisymmetric area tensors (bivectors)

\begin{equation}\label{S-l-l}
\pi_{ab}=\epsilon_{abcd}l^{c}_{1}l^{d}_{2}
\end{equation}
living on triangles spanned by 4-vectors $ l^a_1,~l^a_2$. We use sign
convention for $ \epsilon_{0123}=+1$ and metric signature
(-,~+,~+,~+). The Lagrangian of the continuous time Regge calculus $
L_{\rm Regge}$ includes, among others, the kinetic term,

\begin{equation}\label{kin}
L_{\dot{\Omega}}=\sum_{(ikl)}\pi_{(ikl)}\circ(\bar{\Omega}_{(ikl)}\dot
{\Omega}_{(ikl)}),~~~A\circ B\stackrel{\rm def}{=}{1 \over
2}A^{ab}B_{ab}
\end{equation}
(summed over triangles $ (ikl)$) and the Gauss law one,

\begin{equation}\label{Gauss}
L_h=\sum_{(iklm)}h_{(iklm)}\circ\sum_{{\rm cycle\, perm}\, iklm}

\varepsilon_{(ikl)m}\Omega^{\delta_{(ikl)m}}_{(ikl)}\pi_{(ikl)}
\Omega^{-\delta_{(ikl)m}}_{(ikl)},~~~
\delta \stackrel{\rm def}{=}\frac{1+\varepsilon}{2},\nonumber\\
\end{equation}
summed over tetrahedrons. The $ h_{(iklm)}$ is antisymmetric tensor
Lagrange multiplier, and $ \varepsilon_{(ikl)m}=\pm 1$ is sign
function which specifies for each bivector $ \pi$ choice of one of
two tetrahedrons in the frame of which it is defined.

To write out two else terms in $ L_{\rm Regge}$ we denote by $
n_{ik}dt$ bivector of infinitesimal triangle $ (i^+ik)$. The
contribution of the induced in 3D leaf curvature into Lagrangian
reads

\begin{eqnarray}\label{L-n}
L_n & = & \sum_{ik}|n_{ik}\!|\arcsin\frac{n_{ik}}{|n_{ik}\!|}\circ
R_{ik}\\
&&(R_{ik}=\Omega^{\varepsilon_{ikl_n}}_{i(kl_n\!)}\ldots
\Omega^{\varepsilon_{ikl_1}}_{i(kl_1\!)},~~~ \varepsilon_{ikl_j}\!=
-\varepsilon_{(ikl_j)l_{j-1}}\!=\varepsilon_{(ikl_j)l_{j+1}},~~~|n|\st
ackrel{\rm def}{=}(n\circ n)^{1/2}).\nonumber
\end{eqnarray}
The $ \Omega_{i(kl)}$ is brief notation for connection which in the
full discrete theory is defined on infinitesimal 3-tetrahedron $
(i^-ikl)$ or $ (i^+ikl)$ (not the same as $ \Omega_{(ikl)}$; see
below).

Finally, there are specifical for Regge calculus terms resulting from
cancellations between contributions of closely located leaf and
diagonal triangles inside the infinitesimal 3-prism, see Fig.1.

\begin{eqnarray}
\label{prism}
\begin{picture}(60,130)(50,-15)
\put(-100,90){\rm Fig.1. Infinitesimal 3-prism}
\put (100,-20){\line(0,1){120}}
\put (100,-20){\line(3,2){120}}
\put (100,-20){\line(4,1){160}}
\put (100,20){\line(1,1){120}}
\put (100,20){\line(2,1){160}}
\put (100,20){\line(3,1){120}}
\put (100,20){\line(1,0){160}}
\put (100,100){\line(3,1){120}}
\put (100,100){\line(1,0){160}}
\put (220,60){\line(0,1){80}}
\put (220,60){\line(1,-1){40}}
\put (220,140){\line(1,-1){40}}
\put (220,140){\line(1,-3){40}}
\put (260,20){\line(0,1){80}}
\put (260,95){$~l^{+}$}
\put (220,140){$~k^{+}$}
\put (87,95){$i^{+}$}
\put (92,15){$i$}
\put (220,60){$~k$}
\put (260,15){$~l$}
\put (87,-25){$i^{-}$}
\end{picture}\nonumber
\end{eqnarray}

Each such triangle in the full 4D Regge manifold carries, generally
speaking, finite at $ dt\rightarrow 0$ curvature which is path
ordered product of connection matrices around this triangle. For
example, write out the finite part of curvature matrix $R_{(ikl)}$
if,
e.g., triangle $(ikl)$ is common 2-face of the infinitesimal
tetrahedrons
$(i^{-}ikl)$ and $(ik^{+}kl)$. Finite part is due to connections just
on infinitesimal tetrahedrons: connection on leaf tetrahedron ensure
parallel vector transport at a distance $ O(dt)$ and in the $
dt\rightarrow 0$ limit should be put to be infinitesimal. So we have

\begin{equation}
R_{(ikl)}=\bar{\Omega}_{i(kl)}\Omega_{k(li)}+O(dt)
\end{equation}
Therefore corresponding contribution to action would be $ O(1)$ in
this limit thus giving infinite contribution to Lagrangian. However,
due to eqs. of motion contributions of closely located diagonal and
leaf triangles cancel each other up to desired $ O(dt)$ terms; at the
same time some relation between $ \Omega_{i(kl)}$, $ \Omega_{k(li)}$
and $ \Omega_{l(ik)}$ arises. To see this we make infinitesimal
variations of connections on the infinitesimal tetrahedrons, e.g.

\begin{equation}
\delta\Omega_{k(li)}=w_{k(li)}\Omega_{k(li)}dt,~~~\bar{w}_{k(li)}=
-w_{k(li)}
\end{equation}
at which finite addends to the Lagrangian will arise only from
triangles with finite areas. There are two such terms containing
$\Omega_{k(li)}$ - contributions of $R_{(ikl)}$ and $R_{(ik^{+}l)}$.

Resulting variation of $L$ is linear in $w_{k(li)}$ and leads to a
constraint. Solution of such the constraints for three successive in
$ t$ infinitesimal tetrahedrons takes the form

\begin{equation}
\Omega_{i(kl)}=\Omega_{(ikl)}\exp(\phi_{i(kl)}\pi_{(ikl)}+
\,^{*}\!\phi_{i(kl)}\,^{*}\!\pi_{(ikl)}),~~~^{*}\!A_{ab}\stackrel{\rm
def}{=}{1 \over 2}\epsilon_{abcd}A^{cd}
\end{equation}
\noindent where $\phi_{i(kl)},\,^{*}\!\phi_{i(kl)}$ are parameters.
It is just that SO(3,1) matrix in the RHS denoted $ \Omega_{(ikl)}$
which has already appeared in the Lagrangian terms above.
Corresponding finite contribution into action is proportional to $
\phi_{k(li)}-\phi_{i(kl)}$ and area of the triangle $ (ikl)$; the sum
of such contributions over three successive in $ t$ triangles inside
infinitesimal 3-prism is zero. As for the $ O(dt)$ contribution into
action, it is defined by the differences of areas of these triangles
of the type $ \pi\circ ndt$. This leads to the finite term in the
Lagrangian

\begin{equation}
L_{\phi}=-\sum_{(ikl)}\pi_{(ikl)m}\circ\sum_{{\rm perm}\, ikl}
\varepsilon_{ikl}\phi_{i(kl)}n_{ik(lm)}
\end{equation}
Here more detailed notations $ \pi_{(ikl)m}$ and $ n_{ik(lm)}$ show
that given quantities are defined in the frame of the tetrahedron $
(iklm)$.

The resulting Lagrangian is the sum of the above four terms,

\begin{equation}
L_{\rm Regge}=L_{\dot{\Omega}}+L_h+L_n+L_{\phi}.
\end{equation}
It should be accomplished with the system of bilinear constraints on
bivectors $ \pi,~n$. Some of them state, as in the continuum theory,
that these bivectors should be expressible in terms of edge vectors
in any 4-tetrahedron just as bivectors of continuum theory in terms
of tetrad are. These constraints are of the type $ \pi*\pi,~n*n$ or $
\pi*n$ in the local frame of each tetrahedron where

$$A*B\stackrel{\rm def}{=}{1 \over 4}\epsilon_{abcd}A^{ab}B^{cd}$$

Other constraints special for Regge calculus are continuous time
version of those on scalar products of bivectors $(\cdot\circ\cdot)$
which state independence of these products on the 4-tetrahedron in
the frame of which these are computed. Note that this implies
extension of the set $ \pi,~n$ entering $ L_{\rm Regge}$ (where each
bivector was defined in only one frame) to include each bivector
defined in the frames of {\it all} the 4-tetrahedrons containing
corresponding triangle. To see that these constraints form set
sufficient for that each link would have the length independent on
the 4-tetrahedron where it is found note that each 3-tetrahedron has
6 independent scalar products of it's face bivectors, the same as the
number of it's edges; therefore continuity of the scalar products
when passing between the two 4-tetrahedrons having it as their common
face is sufficient to ensure continuity of edges.

Now in the continuous time on the infinitesimal tetrahedron
continuity condition of scalar products of the type $
\pi\circ\pi,~n\circ n$ or $ \pi\circ n$ does not contain time
derivatives. As for the leaf tetrahedron, it separates two close in
time 4-tetrahedrons; correspondingly, there can be derivatives.
Indeed, given that on the leaf tetrahedron $ (iklm)$ scalar product
of it's two face bivectors $ \pi_{(ikl)m}\circ \pi_{(ikm)l}$ or the
scalar square $ |\pi_{(ikl)m}|^2$ is defined unambiguously at the
time $ t$ we can write out such unambiguity at $ t+dt$ by equating
expressions for this product obtained by two ways. First, we can pass
to each $ t+dt$ triangle, $ (i^+k^+l^+)$ or $ (i^+k^+m^+)$, through
the two intermediate diagonal ones each time writing variation of
triangle bivector $ \delta\pi$ in terms of lateral face bivectors $
ndt$ and calculating corresponding variation of scalar product of the
type $ \pi\circ\delta\pi$. Thus we approach $ t+dt$ triangles from
inside of the infinitesimal 4-prism $ (iklm)$. Second, we can simply
take this product at $ t+dt$ thus approaching from outside of this
prism. Equating both expressions we find the constraints in the form

\begin{eqnarray}
\label{pi-d-pi}
\pi_{(ikl)m}\circ (\dot{\pi}_{(ikl)m}-\sum_{{\rm perm}\,
ikl}\varepsilon_{ikl}n_{ik(lm)}) & = & 0\\

\label{pi-d-pi'}
\pi_{(ikm)l}\circ (\dot{\pi}_{(ikl)m}-\sum_{{\rm perm}\,
ikl}\varepsilon_{ikl}n_{ik(lm)})
+(l\leftrightarrow m) & = & 0
\end{eqnarray}

Central point of the note is that (\ref{pi-d-pi}), (\ref{pi-d-pi'})
are consequences of eqs. of motion and of other constraints without
derivatives. First consider (\ref{pi-d-pi}) for $ m=m_0$ such that $
\pi_{(ikl)m_0}$ is just $ \pi_{(ikl)}$ entering $ L_{\rm Regge}$. The
considered constraint just follows from Regge action upon variation

\begin{equation}
\delta\Omega_{(ikl)}=\Omega_{(ikl)}\pi_{(ikl)}\delta\lambda_{(ikl)},~~
{}~\delta\phi_{(ik)l}=-\delta\lambda_{(ikl)},~~~{\rm
cycle~perm}~i,~k,~l,
\end{equation}
$ \delta\lambda$ being infinitesimal parameters. If $ m\neq m_0$, eq.
(\ref{pi-d-pi}) follows from that already found for $ m_0$ by
differentiating scalar product constraint without derivative, $
|\pi_{(ikl)m}|^2=|\pi_{(ikl)m_0}|^2$. Generally we note that the
number of areas is larger then the number of linklengths. This is
because each triangle has three edges but each edge is shared by no
less then three triangles. This means that validity of all the
relations (\ref{pi-d-pi}) even only for $ m=m_0$ is already
sufficient for unambiguous definition of linklengths on junctions of
4-prisms. Formally, linklengths and therefore scalar products of
different bivectors are expressible in terms of areas by means of all
other kinematical constraints without derivatives; we can
differentiate thus expressed scalar products with the help of
(\ref{pi-d-pi}). Since the relation thus obtained is purely
kinematical one it will be nothing but (\ref{pi-d-pi'}).

Thus, Regge action partially contains kinematical information on
Regge manifold. Important role is played here by terms $ L_\phi$ and
variables $ \phi$ not having continuum analogs.
These variables although entering nonlinearly play the role analogous
to that of Lagrange multipliers at Gauss law enforcing desired
kinematical relation now with time derivatives.

\bigskip
I am grateful to prof. A. Niemi, S. Yngve and personnel of Institute
of Theoretical Physics at Uppsala University for warm hospitality and
support during the work on this paper.


\begin{thebibliography}{99}
\bibitem{Rom}
 J.D.Romano, {\it Gen.Rel.Grav.~}{\bf 25}~(1993)~759
\bibitem{Regge}
 T.Regge, {\it Nuovo Cim.~}{\bf 19}~(1961)~558
\bibitem{Will-rev}
 R.M.Williams and P.A.Tuckey, {\it Class.Quantum Grav.~}{\bf
9}~(1992)~1409
\bibitem{Kha}
 V.Khatsymovsky, {\it Class.Quantum Grav.~}{\bf 6}~(1989)~L249;~{\bf
8}~(1991)~1205
\bibitem{Kha1}
 V.Khatsymovsky, {\it Regge calculus in the canonical form}
Novosibirsk preprint BINP 93-42 (1993)
\end{thebibliography}
\end{document}